%% file: virialc_4.tex
\documentclass[12pt]{article}

\catcode`\@=11

\global\arraycolsep=2pt
\usepackage{amsbsy,amssymb,latexsym,amsfonts,amsmath}
\usepackage{graphicx,color}

\definecolor{refkey}{rgb}{0.9451,0.2706,0.4941}
\definecolor{labelkey}{rgb}{0.2706,0.4941,0.9451}

\newcommand{\B}[1]{{\mathbb #1}}
\newcommand{\C}[1]{{\mathcal #1}}
\newcommand{\F}[1]{{\mathfrak #1}}

\newcommand{\BS}[1]{{\boldsymbol #1}}

\newcommand{\beq}{\begin{equation}}
\newcommand{\eeq}{\end{equation}}
\newcommand{\bea}{\begin{eqnarray}}
\newcommand{\eea}{\end{eqnarray}}
\newcommand{\nn}{\nonumber}

\newcommand{\lish}{\langle\!\langle}
\newcommand{\rish}{\rangle\!\rangle}

\newcommand{\half}{\frac 12}
\newcommand{\third}{\frac 13}
\newcommand{\quarter}{\frac 14}

\newcommand{\eighth}{\frac 18}

\newcommand{\Slash}[1]{{\ooalign{\hfil#1\hfil\crcr\raise.167ex\hbox{/}}}}

\begin{document}
\begin{flushright}
\parbox{4.2cm}
{IPMU14-0067}
\end{flushright}

\vspace*{0.7cm}

\begin{center}
{ \Large\bf 
Improvement of energy-momentum tensor and\\
\vspace{4pt}
non-Gaussianities in holographic cosmology
}
\vspace*{1.5cm}\\
{Shinsuke Kawai}${}^a$
and
{Yu Nakayama}${}^{b,c}$
\end{center}
\begin{center}
${}^a${\it Department of Physics, Sungkyunkwan University,
Suwon 440-746, Republic of Korea} \\
\vspace{7pt}
${}^b${\it California Institute of Technology,
452-48, Pasadena, California 91125, USA} \\
\vspace{7pt}
${}^c${\it Kavli Institute for the Physics and Mathematics of the Universe (WPI), \\
Todai Institutes for Advanced Study, Kashiwa, Chiba 277-8583, Japan}
\vspace{30pt}
\end{center}

\begin{abstract}
In holographic models of cosmology based on the (A)dS/CFT correspondence, conformal symmetry is implicit in the dual description of the Universe.
Generically, however, one cannot expect the (broken) conformal invariance in the cosmic fluctuations as only the scale invariance is manifest in experiments.
Also, in order for the prediction of the holographic models to make sense, the conformal symmetry needs to be broken as the scalar mode of the metric fluctuations becomes pure gauge in the conformal limit.
We discuss the improvement ambiguity of the energy-momentum tensor in this context and 
construct a holographic model of the Universe that preserves the scale invariance but not necessarily the full conformal invariance.
Our sample computation using a weakly coupled dual field theory shows that the orthogonal type of non-Gaussianity is present over and above the equilateral type.
The improvement ambiguity corresponds to the choice of the energy momentum tensor
that will couple to our particle physics sector after inflation.
Our results show that the holographic prediction of the cosmological parameters crucially depends on such a choice.
\end{abstract}


\thispagestyle{empty} 

\setcounter{page}{0}

\newpage

\tableofcontents

\section{Introduction}
\label{sec:intro}
While inflationary cosmology is extremely successful in explaining the initial conditions of the Big Bang Universe, there are unsolved issues.
One issue is that its particle physics origin, i.e. which field at which energy scale is responsible for the realization of inflation, is not known as of today.
Another, perhaps a more fundamental, problem is the validity of the use of effective field theory methods:
the beginning of inflation often corresponds to an energy scale close to Planckian,
and generation of sufficient e-folding in generic models  requires super-Planckian excursion of the inflaton field.
These features typically suggest that quantum gravity effects are potentially important.
While we do not have a fully satisfactory framework to discuss these issues at the moment, we could expect 
that technologies developed in the gauge/gravity correspondence may give us insights.

Recently there has been much interest in constructing holographic models of the early Universe
\cite{Maldacena:2002vr,
Larsen:2002et,Larsen:2003pf,Larsen:2004kf,Seery:2006tq,McFadden:2009fg,McFadden:2010na}.
The idea behind is the de Sitter/conformal field theory (dS/CFT) correspondence
\cite{Witten:2001kn,Strominger:2001pn,Maldacena:2002vr} which states
that the wave function of the $(d+1)$-dimensional de Sitter space is equivalent to the partition function of a $d$-dimensional conformal field theory.
Although the dS/CFT correspondence is not as rigorous as the AdS/CFT correspondence, 
and the holographic approach to cosmology is not mature enough to provide a reliable transition mechanism from the de Sitter expansion to the radiation dominated era, the idea of holography comes with a concrete prescription for computing the spectra of the primordial fluctuations which is
based on an analytic continuation of the AdS/CFT dictionary
\cite{Maldacena:1997re,Gubser:1998bc,Witten:1998qj}.
Encouraging results have been reported in 
\cite{McFadden:2009fg,McFadden:2010na,McFadden:2010vh,Antoniadis:2011ib,Bzowski:2011ab,McFadden:2011kk}
that observationally viable primordial spectra are obtained by assuming specific examples 
of dual quantum field theory.
In this way, holography may provide a UV complete description of the early Universe that includes a mechanism to generate primordial fluctuations.
Notably, these models can be in principle tested by future experiments through their
prediction of cosmological observables.

The basis of the holographic interpretation of the early Universe is the approximate de Sitter symmetry.
In fact, the description of the primordial spectra in the light of the de Sitter symmetry is itself an interesting subject, and is also of practical use, irrespective of whether the fluctuations are generated holographically or by inflation
\cite{Maldacena:2002vr,Larsen:2002et,Larsen:2003pf,Larsen:2004kf,Maldacena:2011nz,Creminelli:2011mw}.
It is believed that the scale invariance of the cosmic microwave background (CMB) spectrum is a
consequence of the de Sitter expansion in the early Universe.
More to the point, the scale invariance is associated with the dilatation symmetry within the three
dimensional conformal group.
The isometry of the $3+1$ dimensional de Sitter space is $SO(4,1)$, generated by
$J_{ab}$ ($a,b = 0,1, 2, 3, 4$) satisfying the algebra
\beq
[J_{ab}, J_{cd}]=\eta_{ac}J_{bd}+\eta_{bd}J_{ac}-\eta_{ad}J_{bc}-\eta_{bc}J_{ad},
\quad
\eta_{ab}={\rm diag}(-1,1,1,1,1).
\eeq
The de Sitter group is isomorphic to the three dimensional conformal group that consists of 
translation (generated by $P_i$), rotation ($L_{ij}$), dilatation ($D$) and the special 
conformal transformation (SCT) ($K_i$, $i, j=1,2,3$), into which the $SO(4,1)$ generators are decomposed,
\beq
J_{ij}=L_{ij},\quad
J_{04}=D,\quad
J_{0i}=\half (P_i-K_i),\quad
J_{4i}=\half (P_i+K_i).
\eeq
In the inflationary patch (the planar coordinates) of the de Sitter space 
\beq
ds^2=\frac{-d\tau^2+d{x^i}^2}{H^2\tau^2},
\label{eqn:dS_metric}
\eeq
where $H$ is the Hubble parameter and $\tau$ the conformal time, 
the translation and the rotation are symmetries of the constant time hypersurface, 
reflecting the homogeneity and isotropy of the space (which the inflation is suppose to achieve in the end \cite{Wald:1983ky}).
The dilatation corresponds to simultaneous scaling of the space and time
$(\tau,x^i)\to (\lambda\tau, \lambda x^i)$,
which, combined with the (approximate) shift symmetry of the inflaton field 
$\varphi\to\varphi+{\rm const.}$, is the origin of the scale invariance of the fluctuation 
spectrum.
The SCT corresponds to the isometries of the de Sitter space
\beq
\tau\to\tau+2 ({\BS b}\cdot {\BS x}) \tau, \quad
x^i\to x^i+(\tau^2-{\BS x}^2) b^i+2 ({\BS b}\cdot {\BS x}) x^i,
\label{eqn:dSSCT}
\eeq
where $b^i={\BS b}$ is an infinitesimal 3-vector.
The implication of this symmetry in the case of inflationary cosmology is not as immediate as the other symmetries, as it acts
nonlinearly on the spacetime coordinates and maps a constant inflaton 
hypersurface to another hypersurface on which the inflaton is not constant any more.
Note that the isometries (\ref{eqn:dSSCT}) coincide with the SCT of the three dimensional
conformal group at late times $\tau\to 0$.
Here, $-\infty<\tau<0$.

In slow roll inflation, the dilatation and SCT symmetries are broken by the inflaton dynamics,
whereas the translation and rotation remain (asymptotically) exact symmetries of the background spacetime.
As long as the slow roll conditions are satisfied, nevertheless, the full conformal symmetry stays as a good approximate symmetry of the inflationary spacetime.
The implication of this approximate conformal symmetry on the primordial fluctuation spectra has been 
a focus of much attention 
\cite{Maldacena:2011nz,Creminelli:2011mw,Assassi:2012zq,Assassi:2012et,Kehagias:2012pd,Kehagias:2012td,Hinterbichler:2011qk,Hinterbichler:2012nm,Hinterbichler:2012mv,Mata:2012bx,Ghosh:2014kba,Garriga:2013rpa,Garriga:2014ema} . 
It has been pointed out that the correlation functions of the primordial gravitational waves
\cite{Maldacena:2011nz} and scalar fluctuations with negligible coupling to the inflaton (such as in the case of curvatons) \cite{Creminelli:2011mw} are constrained by the broken conformal symmetry.
For fluctuations sourced by the inflaton, one cannot expect the spectra to enjoy the full conformal symmetry, as the dynamics of the inflaton breaks the symmetry explicitly.
See however \cite{Creminelli:2012ed,Schalm:2012pi,Hinterbichler:2013dpa,Berezhiani:2013ewa}  for the discussion on the consistency relations obtained by
the remnant of the de Sitter symmetry.

The three dimensional conformal symmetry is a much larger symmetry than the Euclidean (or Poincar\'{e}) $+$ scaling symmetry, as the former includes 10 parameters (associated with $P_i$, $L_{ij}$, $D$, $K_{i}$) 
whereas the latter only 7 ($P_i$, $L_{ij}$, $D$).
In quantum field theory the distinction is rather subtle, as it is known that in {\em reasonable} unitary field theories, the 
Poincar\'{e} $+$ scaling symmetry is enhanced to the conformal symmetry (see \cite{Nakayama:2013is} for a review).\footnote{The situation in  Euclidean field theories is much more subtle because the unitarity is replaced by reflection positivity, and there is no fundamental reason why it must be satisfied. See e.g. \cite{Riva:2005gd} for a physical counterexample.}
This enhancement is related to the existence of the so-called improvement ambiguity of the energy-momentum tensor.
In dS/CFT correspondence, the field theory dual to a de Sitter space (if it exists) is not necessarily unitary (more precisely reflection positive), and hence it is particularly important to make distinction between scale invariance and conformal invariance.
In this paper we revisit the holographic models of cosmology 
\cite{McFadden:2009fg,McFadden:2010na,McFadden:2010vh,Bzowski:2011ab,McFadden:2011kk} from the viewpoint of scaling/conformal symmetry, 
focusing on the role played by the improvement of the energy-momentum tensor.

The following sections are organised as follows.
In the next section we start by reviewing the improvement of the energy-momentum tensor 
in general contexts in quantum field theory.
We discuss the holographic interpretation of the improvement of the energy-momentum tensor in the weakly coupled gravity regime in Sec.\ref{sec:localRG}.
An example of the weakly coupled three-dimensional field theory is studied in Sec.\ref{sec:3dQFT}, with computation of correlators including the improvement ambiguity.
The implications of this computation on the cosmological observables are discussed in Sec.\ref{sec:observables}.
We conclude in Sec.\ref{sec:concl} with discussions. 
In three Appendices we summarize some technicalities and conventions that we use in the paper.

\section{Improvement of the energy-momentum tensor}
\label{sec:improvement}
Let us consider a local quantum field theory in $d$ dimensions ($d=3$ will be our main focus) with Euclidean (or Poincar\'e) invariance. 
The energy-momentum tensor can be defined as the canonically conserved Noether current associated with translation.
It can be made symmetric if rotational (or Lorentz) symmetry is present, through the Belinfante prescription.
Hence, in the presence of the full Euclidean (or Poincar\'{e}) symmetry, the energy-momentum tensor $T_{ij}$ can be defined, 
and made symmetric.
The energy-momentum tensor so defined is not unique, due to the so-called improvement ambiguity
\begin{align}
\tilde{T}_{ij} &= T_{ij} 
+ \eta \left(\partial_i \partial_\ell L^{\ell}_{\  j} 
+ \partial_j \partial_\ell L^{\ell}_{\ i} 
- \Box L_{ij} - \delta_{ij} \partial_\ell\partial_k L^{\ell k} \right)  
+ \kappa (\delta_{ij} \Box L^{\ell}_{\ \ell} - \partial_i \partial_j L^{\ell}_{\ \ell} ),
\label{eqn:improvement}
\end{align}
with $L_{ij}$ an arbitrary symmetric tensor operator. Here $\delta_{ij}$ is $d$-dimensional Euclidean (or Lorentzian) flat metric.
The extra terms proportional to $\eta$ and $\kappa$ on the right hand side do not contribute to the Euclidean (or Poincar\'e) charges and they are conserved by themselves without using dynamical information.
A typical example of such an improvement is obtained by the curvature coupling of a scalar 
$\delta S = \int d^dx \sqrt{|g|} \xi R \phi^2$, leading to $L_{ij} = \xi \delta_{ij} \phi^2$ above.
Here, we use the fact that the symmetric energy-momentum tensor can be written as
\begin{align}
 T_{ij} = \left. \frac{2}{\sqrt{|g|}} \frac{\delta S}{\delta g^{ij}} \right |_{g_{ij} = \delta_{ij}}
 \label{eqn:Tmunu}
\end{align}
from the curved space action.
More generally, the improvement (\ref{eqn:improvement}) corresponds to the modification of the gravitational coupling as
\begin{align}
\delta S = \int d^dx \sqrt{|g|} (-\eta R_{ij} L^{ij} + \frac{\kappa}{2} R L^\ell_{ \ \ell}),
\end{align} 
where $R_{ij}$ is the $d$-dimensional Ricci tensor and $R$ is the scalar curvature.

The response of a quantum field theory to the conformal transformation is governed by the trace of the 
energy-momentum tensor.
The theory becomes conformal invariant in flat spacetime when the trace of the energy-momentum tensor vanishes,
\begin{align}
T^{i}_{\ i} = 0 .
\end{align}
Given the ambiguity (\ref{eqn:improvement}), the theory is said to be improved to be conformal invariant when the trace of the energy-momentum tensor in the flat spacetime takes the form
\begin{align}
T^{\ell}_{\ \ell} = \partial_i \partial_j L^{ij},
\end{align}
with $\eta = \frac{1}{d-2}$ and $\kappa = \frac{1}{(d-2)(d-1)}$. 
It should be emphasized that while the improvement does change the coupling to the background metric, it does not affect the physics in the limit of flat spacetime as long as the operator 
$L_{ij}$ is a well-defined local operator.

In a power-counting renormalizable field theory, the trace of the energy-momentum tensor in flat spacetime generically takes the form\footnote{Due to the unitarity constraint, we do not have higher tensor operators on the right hand side of this equality near the conformal fixed point as in perturbation theory around a Gaussian fixed point. In particular the term like $\partial_{i}\partial_{j} L^{ij}$ is only allowed when $L_{ij}$ is proportional to the metric $L_{ij} = \delta_{ij} L$.}
\begin{align}
T^{i}_{\ i} = \beta^I O_I + \partial^i J_i + \kappa^\alpha \Box O_\alpha \  \label{traceid}
\end{align}
up to equations of motion.
The equation \eqref{traceid} is known as the trace identity, and it gives the local Callan-Symanzik equation.
In many situations (e.g. perturbation theory around a Gaussian fixed point), it is possible to remove the 
$\partial^i J_i$ term by using operator identities. 
The argument above also suggests that the term $\kappa^\alpha \Box O_\alpha$ can be removed by the
improvement transformation if we are allowed to change the gravitational coupling. 
These total derivative terms do not contribute to the global scale transformation, as they vanish 
upon integration by parts for the constant Weyl transformation.

In holographic cosmology, the trace identity \eqref{traceid} is interpreted as the Hamiltonian constraint of the dual gravitational system.
The first term in \eqref{traceid} may be understood as the running of the 
inflaton that induces deviation from the approximate de Sitter spacetime. In a certain limiting regime, the connection between the slow roll inflation and the conformal perturbation theory is understood \cite{Larsen:2003pf,Larsen:2004kf,Seery:2006tq,vanderSchaar:2003sz,Bzowski:2012ih,Kiritsis:2013gia,Kol:2013msa,Bourdier:2013axa}. 
The contribution from the second term $\partial_i J^i$ is related to the effect of dynamical vector fields during inflation, which will be discussed somewhere else.
Our main focus of this paper is the role of the improvement term $\kappa^\alpha \Box O_\alpha$ and its implication in holographic cosmology.

\section{Interpretation of improvement in weakly coupled bulk}
\label{sec:localRG}

The holographic models of cosmology is based on holographic renormalization group flows in the dS/CFT correspondence, which is a particular analytic continuation of the more robust 
AdS/CFT correspondence and its holographic renormalization flow. Our discussions in the following may be applied to the AdS cases with minor modifications.

The holographic cosmology is based on the idea that the wave function of the universe is obtained by the (analytic continuation of) the generating functional of the dual quantum field theory similarly to the case in AdS spacetime \`{a} la GKP-W prescription
\cite{Gubser:1998bc,Witten:1998qj}. The quantum fluctuations of the universe encoded in the wave function will give rise to the primordial CMB fluctuations after inflation.

While we will focus on the weakly coupled dual field theory computation in the following sections (i.e. strongly coupled bulk gravity computation), in this section we would like to give an interpretation of the improvement ambiguities of the dual field theory from the weakly coupled bulk description. The analysis here serves as a bridge between the holographic computation and the slow roll inflation in terms of the effective field theory in the bulk.

To begin with, let us consider the situation at the (conformal) fixed point described by the de Sitter metric in the Poincar\'e patch
\begin{align}
ds^2 = G_{\mu\nu} dx^\mu dx^\nu =\frac{-d\tau^2+d{x^i}^2}{H^2\tau^2},
\end{align}
where $i= 1,\cdots, d$.
Obviously the scale transformation is generated by
\begin{align}
\tau &\to \lambda \tau \cr
x^i &\to \lambda x^i , 
\end{align}
and we regard the conformal time $\tau$ as the renormalization group scale $\mu$ of the dual field theory: $H \tau \sim \log \mu$.
We typically assume that the gravitational theory is $(d+1)$-dimensional diffeomorphism invariant and there is no other field condensation so that the dual theory is essentially at the conformal fixed point.\footnote{If we lift these assumptions, it is possible that the fixed point is only scale invariant but non-conformal invariant \cite{Nakayama:2012sn}.}

The basis of the dS/CFT correspondence is the analytic continuation of the GKP-W prescription in AdS/CFT correspondence. In the weakly coupled bulk regime, we identify the wave function of the universe as the Schwinger functional of the dual conformal field theory
\beq
\Psi_{\rm dS}[g_{ij}(x), g^I(x)]=Z_{\rm CFT}[g_{ij}(x), g^I(x)],
\label{eqn:dSCFT_rel}
\eeq
where on the left hand side the $d$-dimensional metric $g_{ij}(x)$ and the scalar source field $g^I(x)$ fix the boundary values of the bulk metric $G_{\mu\nu}$ and the bulk scalar
$\Phi^I$ at the future boundary.
See section \ref{sec:observables} for details of its relation to the primordial fluctuations of the Universe.

From the dual field theory viewpoint, the above assumption of the de Sitter invariance means that we have eliminated the first two terms in the trace identity \eqref{traceid}. The scalar beta functions $\beta^I$ are absent because the scalar fields $\Phi^I$
whose boundary value is fixed by $g^I$ do not evolve along the time direction $\tau$ (i.e. the renormalization group direction) and the theory is scale invariant. The contribution from $\partial_i J^i$ is given by the non-trivial vector fields in the bulk, which we have also assumed not to exist. What we would like to study in the following is the effect of the last improvement term, which can still exist even in the de Sitter background, from the holographic viewpoint.

In the weakly coupled gravity regime, the implementation of the improvement of the energy-momentum tensor in holography goes as follows.\footnote{Again, due to the unitarity reasoning, we focus on the improvement from the scalar operators only.}
We assume that the dual field theory at the conformal fixed point contains a scalar operator $O_\alpha$ with conformal dimension $\Delta = d-2$. 
Correspondingly we have a scalar field $\Phi^\alpha$ in the bulk whose mass is given precisely by the conformal coupling in $d+1$ dimensions. Schematically the scalar action is given by 
\begin{align}
S = \int d^{d+1} x \sqrt{|G|} \left(G^{\mu\nu}\partial_\mu \Phi^\alpha \partial_\nu \Phi^\alpha + \frac{d-1}{4d} R (\Phi^\alpha)^2  + \text{interactions} \right) .
\end{align}
There is no particular reason why the mass must be given by the coupling to the bulk curvature. The origin of the conformal mass for the scalar field $\Phi^\alpha$ only affects the higher order correlation functions that we will not dwell on.

The analytic continuation of the GKP-W prescription for the wave function of the universe suggests that we identify the field theory Schwinger functional with the source 
$g_{ij}(x)$ for the energy momentum tensor and the source 
$m^\alpha(x)$ 
for the scalar operator $O_\alpha$ as the wave function of the universe with the boundary condition
\begin{align}
G_{ij} &\sim \frac{g_{ij} (x)}{H^2\tau^2} + \cdots \cr
\Phi^\alpha & \sim \frac{m^\alpha(x)} {H^2\tau^2} + \cdots
\end{align}
as we approach the future boundary $\tau \to 0$. 
The asymptotics for the scalar field is dictated by the requirement that the corresponding operator has the conformal dimension $\Delta = d-2$.

At the superficial level, the holographic implementation of the improvement of the energy-momentum tensor of the dual field theory is simple: we consider that the source field for the operator $O_\alpha$ having dimension $\Delta = d-2$ is mixed with the $d$-dimensional curvature term:
\begin{align}
\Phi^\alpha \to \tilde{\Phi}^\alpha & \sim \frac{m^\alpha(x) + \delta\xi^\alpha R^{(d)}(x) }{H^2\tau^2} + \cdots  \label{changeb}
\end{align}
with arbitrary parameters $\delta\xi^\alpha$,
where $R^{(d)}(x)$ is the scalar curvature constructed out of the $d$-dimensional source metric $g_{ij}(x)$.

The boundary condition such as \eqref{changeb} may be cast into a more standard form by performing the field redefinition in the $(d+1)$-dimensional bulk. We can simply replace the bulk scalar field $\Phi^\alpha$ with $\tilde{\Phi}^\alpha = \Phi^\alpha + \delta\xi^\alpha R$ where $R$ is the $(d+1)$-dimensional scalar curvature (rather than the $d$-dimensional curvature which is $R^{(d)}$). Then we can simply use the standard GKP-W boundary conditions and compute the wave function of the universe, whose result is the same as the improvement of the energy-momentum tensor discussed above.

The shift $\Phi^\alpha \to \tilde{\Phi}^\alpha =  \Phi^\alpha + \delta\xi^\alpha R$ generically induces effective linear coupling $\delta \xi^\alpha \tilde{\Phi}^\alpha R$ in the bulk action, and therefore the trace of the energy-momentum tensor, which couples to the scalar mode of the metric fluctuations in the bulk, becomes non-zero even in the de Sitter background.  
As expected from the field theory discussions (see also Appendix A), the scalar mode
fluctuations of the new metric after the field redefinition originate from the fluctuations of $\Phi^\alpha$ which showed the conformal invariant fluctuations before the field redefinition.\footnote{As another check, one may also compute the holographic Weyl anomaly in $d=4$. It can be easily shown that in addition to the standard $c$ anomaly (Weyl squared) and $a$ anomaly (Euler density), we have the additional term $\delta \xi^\alpha R^2$.  This term does not satisfy the Wess-Zumino consistency condition if the theory is conformal invariant, but with the improvement, the $R^2$ anomaly is allowed, and the appearance of the term proportional to $\delta \xi^\alpha$ is of course consistent with what was obtained from the local renormalization group analysis.}

This prescription is consistent with the local renormalization group analysis of the renormalized Schwinger functional with the so-called class 2 ambiguity discussed in \cite{Nakayama:2013wda}, which can be generalized to the  generic power-counting renormalization beyond conformal fixed points. We will further discuss the connection to the local renormalization group approach in Appendix A, where the non-trivial running of the scalar function (e.g. as in the case with the inflaton) is also introduced.

What we have discussed so far is essentially renaming of the bulk fields, and now we would like to 
explain why such renaming will affect observables, such as the CMB spectrum, in holographic 
cosmology. 
The crucial question here is which operator actually couples to our particle physics sector (and dark matter) after the end of inflation. 

A typical assumption made in the standard slow roll inflationary cosmology is that the inflaton couples to the particle physics operators either directly or indirectly, and its eventual decay into the Standard Model particles thermalizes the Universe.
The fluctuations of the ``metric" (scalar mode of which is related to the inflaton fluctuations by the Hamiltonian constraint) generated quantum mechanically during inflation 
are streched to superhorizon scales and become the primordial fluctuations of the Universe.
Once they are out of the horizon scale 
their amplitude will not change; in particular, the spectra are not affected by details of the reheating mechanism.
If multiple light degrees of freedom are present during inflation, there are
isocurvature modes which may be converted into the density fluctuations during or after inflation, as in the case of the curvaton scenario.
Moreover, density fluctuations may be generated during the reheating, for example, by the modulated reheating mechanism. 
The tacit assumption here is that the ``metric" during inflation and the ``metric" that couples to the Standard Model particles after inflation share the same seed of fluctuations throughout the reheating process.

In the holographic cosmological model, an analogue of the curvaton mechanism
is such that an operator whose beta function is small will eventually couple to the particle physics sector, giving fluctuations of the particle physics sector independently of the metric fluctuations coming from the dual energy-momentum tensor. The wave function of the universe computed from the GKP-W prescription determines the fluctuations for both metric and additional scalars; by specifying how it couples to the particle physics the subsequent evolution of the Universe should be determined.

In addition to these light fields, we have introduced the conformal scalar $\Phi^\alpha$ in the holographic model corresponding to possible improvement of the energy-momentum tensor. 
This could break the tacit assumption mentioned above.
We argued that the mixing between the scalar and the ``metric" corresponds to the improvement in the dual quantum field theory. The point is that the metric fluctuations $h_{\mu\nu}$ during inflation (the dS/CFT era),
\begin{align}
 \delta S = \int d^{d+1} x h_{\mu\nu} T^{\mu\nu} + \Phi^\alpha O_\alpha \label{original}
\end{align}
may be different from $\tilde h_{\mu\nu}$ which couple to the Standard Model energy-momentum tensor after inflation
\begin{align}
\delta S  = \int d^{d+1} x \tilde{h}_{\mu\nu} T^{\mu\nu}_{(SM)} + \tilde{\Phi}^\alpha O_{\alpha(SM)} ,
\end{align}
which we observe for example in the CMB.
The existence of the improvement ambiguity implies that there is arbitrariness 
in choosing the metric that seeds observed fluctuations.\footnote{
At this point, the discussion becomes somewhat speculative as no concrete reheating mechanism nor 
concrete string-theoretical implementation has been worked out in the holographic model. 
The distinction between $h_{\mu\nu}$ and $\tilde h_{\mu\nu}$ may seem analogous to the choice of the Jordan frame or the Einstein frame, but the origin is different.
We discuss the holographic interpretation of the frame choice in Appendix A. Note that in the latter case the observables do not depend on frames (see for example a review \cite{Yamaguchi:2011kg}), but with our improvement ambiguities, the prediction depends on the choice of the energy-momentum tensor.}
Besides, the particle physics energy-momentum tensor $T^{\mu\nu}_{(SM)}$ itself involves
some arbitrariness (e.g. nonminimal curvature coupling to the Higgs field).
Depending on the mechanism of inflation and how it ends, the moduli stabilization for instance may suggest that
\begin{align}
h_{\mu\nu} &\to \tilde{h}_{\mu\nu} = h_{\mu\nu} \cr
{\Phi}^\alpha &\to \tilde{\Phi} = {\Phi}^\alpha + \delta \xi^\alpha {R}^{(d)} 
\end{align}
be the natural choice of the field variables as $\tau \to 0$.
In this case we will be interested in the fluctuations of $\tilde{h}_{\mu\nu}$, not of $h_{\mu\nu}$,
in our holographic model of cosmology.
Then the fluctuations of the metric that couples to our particle physics fields must be given by the correlation functions of the improved energy-momentum tensor, rather than the original one in \eqref{original}.
Actually, there is no preferred choice of the energy-momentum tensor here without specifying
an explicit model of how the inflation ends. 
What we only know is that the dual theory is not manifestly conformal invariant if the chosen energy-momentum tensor is not traceless (even though it may be improved to be traceless). In the following sections, we perform sample computation of the cosmic fluctuations from a improved dual quantum field theory, to see how the violation of the full de Sitter invariance while preserving the scale invariance affects the cosmological observables.

The above argument is also valid in AdS/CFT with holographic renormalization group. A priori, we do not know how the operator mixes under a renormalization group flow. After solving the bulk equations of motion, one may want to reexamine the boundary conditions for the GKP-W partition function so that the particular fluctuation of the UV boundary gives the desired energy-momentum tensor of the IR theory. Typically, this requires non-trivial mixing of the metric and other scaler fields in the UV limit (see Appendix A for more details). From the dual field theory viewpoint, this corresponds to the choice of $\delta \xi^\alpha$ in \eqref{changeb}, where one ``natural" choice in the UV theory (e.g. $\delta \xi^\alpha = 0$) may or may not be the desired one in the IR. 
One simple example of such a situation is the energy-momentum tensor of a goldstone boson. 
A goldstone boson has a natural shift symmetry, so the curvature coupling $\xi R \phi^2$ is unnatural and the theory is not manifestly conformal invariant and the natural choice of the energy-momentum tensor is an unimproved one. If the UV theory is conformal, the improved traceless energy-momentum tensor may be the natural choice, but this choice would not be natural in the IR where the goldstone boson prefers the unimproved energy-momentum tensor with the manifest shift symmetry.

\section{Dual field theory computation in weakly coupled limit}
\label{sec:3dQFT}
In this section we provide a simple quantum field theory computation for illustrating the effects of 
energy-momentum tensor improvement on correlation functions.
Our example is the three dimensional scalar field theory coupled to the background curvature, with 
the action
\begin{align}
 S = \half \int d^3x \sqrt{g} \left(g^{ij} \partial_i \phi \partial_j \phi + \xi R \phi^2 \right),
 \label{eqn:BosonicAction}
\end{align}
where $i,j=1,2,3$.
It is understood that the space is Euclidean and the flat limit $g_{ij}\to\delta_{ij}$ is taken 
in the end.
As mentioned in Section \ref{sec:improvement}, the curvature coupling leads to an improvement term
in the energy-momentum tensor.
This model (\ref{eqn:BosonicAction}) is a generalization of the model studied in
\cite{McFadden:2009fg,McFadden:2010na,McFadden:2010vh,Bzowski:2011ab},
in which the minimally coupled ($\xi=0$) and conformally coupled ($\xi=\eighth$) scalars, as well as
a gauge field and fermions are considered.

It is straightforward to check that the action (\ref{eqn:BosonicAction}) gives rise to the flat space-time
energy-momentum tensor
\beq
T_{ij}(x)=\partial_i\phi\partial_j\phi
-\half \delta_{ij}(\partial\phi)^2
+\xi(\delta_{ij}\Box-\partial_i\partial_j)\phi^2 ,
\label{eqn:Tij1}
\eeq
where the last term is the improvement term.
On contraction, the trace of the energy-momentum tensor is
\begin{align}
T\equiv T^{i}_{\ i} = -\frac{1}{2} (\partial\phi)^2 
+ 2\xi (\Box \phi^2),
\end{align}
which turns out to be proportional to the equations of motion (and hence vanishes on-shell) at $\xi = \frac{1}{8}$. 
Since we will be interested in application to holographic cosmology below where
semi-local terms in the correlation functions will be important, we keep track of the use of the equations of motion.

According to the dictionary of (A)dS/CFT the metric perturbations in the bulk act as the 
source of the boundary energy-momentum tensor.
Thus, the holographic computation of the cosmological metric fluctuations boils down  to 
computation of the correlation functions of the boundary energy-momentum tensor.
For computing three point (and higher) correlation functions it is also necessary to keep
track of the semi-local contact terms.
We shall be interested in the scalar (density) and tensor (gravitational wave) power spectra
and the scalar bispectrum in this paper.
We define an operator \cite{McFadden:2010vh,Bzowski:2011ab}
\bea
\Upsilon_{ijk\ell}(x,y)&\equiv&\frac{\delta T_{ij}(x)}{\delta g^{k\ell}(y)}\nn\\
&=&\frac{2}{\sqrt{g(x)}}\frac{\delta^2 S}{\delta g^{ij}(x)\delta g^{k\ell}(y)}
+\half T_{ij}(x) g_{k\ell}(x)\delta(x-y).
\eea
For our purposes it is sufficient to know 
$\Upsilon(x,y)\equiv \delta^{ij}\delta^{k\ell} \Upsilon_{ijk\ell} (x,y)|_{g_{ij} = \delta_{ij}}$.
One way of computing this is to consider the response of the action (\ref{eqn:BosonicAction})
under the 
Weyl transformation $g^{ij}\to \widehat g^{ij} = e^{2\sigma} g^{ij}$.
From the functional expansion up to the second order
\begin{align}
S[\widehat g_{ij}]
=S[g_{ij}]
&+\int d^3x \sqrt{g} \left(\sigma T^{i}_{\ i} -\frac{1}{2}\sigma^2 T^{i}_{\ i} \right)\cr
&+\int d^3x \sqrt{g(x)} \int d^3y \sqrt{g(y)} \sigma(x) \sigma(y) \Upsilon_{ijk\ell}(x,y) g^{ij} g^{k\ell}\cdots,
\end{align}
the explicit form of $\Upsilon$ is found to be
\begin{align}
\Upsilon(x,y) 
= -\xi \frac{\partial}{\partial x^i}\left( (\phi(x)^2)\frac{\partial}{\partial x^i} \delta (x-y) \right) .
\label{eqn:Upsilon}
\end{align}

Correlators of $T_{ij}$ and $\Upsilon$ are obtained by straightforward field theory computations, namely, expressing each term of (\ref{eqn:Tij1}) (\ref{eqn:Upsilon}) as a
normal ordered product of the free fields and then performing Feynman integrals. 
The intermediate computations are given in Appendix \ref{sec:FTcomp}.
Note that the theory is super-renormalizable and after regularization the results are finite.
In momentum space, the two point function of the energy-momentum tensor is\footnote{
Momentum conservation restricts ${\BS q}_1,\cdots,{\BS q}_n$ in 
$\lish T_{ij}({\BS q}_1)\cdots T_{k\ell}({\BS q}_n)\rish$ to form an $n$-sided polygon,
and due to rotational invariance the correlator depends only on the magnitude $q_i\equiv |{\BS q}_i|$ of the momenta. See also Appendix \ref{sec:NG}.
}
\begin{align}
&\langle T_{ij}({\BS q}_1) T_{k\ell}({\BS q}_2)\rangle 
=(2\pi)^3\delta^3({\BS q}_1+{\BS q}_2)\lish T_{ij}({\BS q}_1) T_{k\ell}(-{\BS q}_1)\rish,
\cr 
&\lish T_{ij}({\BS q})T_{k\ell}(-{\BS q})\rish
=\Big\{
(3-32\xi+128\xi^2) q_i q_j q_k q_\ell\cr 
&\qquad-q^2\Big[
(1-32\xi+128\xi^2)(g_{ij} q_k q_\ell+g_{k\ell}q_i q_j)
+g_{ik} q_j q_\ell
+g_{i\ell} q_j q_k
+g_{jk}q_i q_\ell
+g_{j\ell}q_i q_k
\Big]\cr 
&\qquad+q^4\big[(1-32\xi+128\xi^2)g_{ij}g_{k\ell}
+g_{ik}g_{j\ell}
+g_{i\ell}g_{jk}\big]
\Big\}\frac{1}{512 q},
\label{eqn:TijTkl_1}
\end{align}
where $q\equiv|{\BS q}|=(\delta^{ij}q_i q_j)^{1/2}$.
Here and below we shall use the double chevrons $\lish\cdots\rish$ to indicate omission of the factor
$(2\pi)^3\delta^3(\sum_i q_i)$ arising from momentum conservation.
Contracting with $\delta^{ij}\delta^{k\ell}$ we obtain
\begin{align}
&\lish T({q}) T(-{q})\rish=\left(\xi-\eighth\right)^2 q^3.
\end{align}
The three point correlator of the trace of the energy-momentum tensor is
(with the same notation as above)
\bea
\lish T({q}_1)T({q}_2)T({q}_3)\rish
&=&-8\left(\xi-\eighth\right)\Big\{
\left(\xi-\eighth\right)^2 q_1q_2q_3
-\frac{1}{128}(q_1+q_2+q_3)(q_1^2+q_2^2+q_3^2)\nn\\
&&\qquad+\eighth\xi\left(q_1q_2^2+q_1^2q_2+q_2q_3^2+q_2^2q_3+q_3q_1^2+q_3^2q_1\right)
\Big\},
\label{eqn:TTT}
\eea
and the cross correlator of the trace of the energy-momentun tensor and the semi-local 
$\Upsilon(x,y)$ operator is
\beq
\lish T({q}_1)\Upsilon({q}_2, {q}_3)\rish
=\quarter\xi\left(\xi-\eighth\right)\left[q_1(q_1^2-q_2^2-q_3^2)\right].
\label{eqn:TUpsilon}
\eeq

\section{Cosmological observables}
\label{sec:observables}

Let us now discuss the holographic prediction for cosmological observables.
The discussion below generalises the model studied in \cite{McFadden:2009fg,McFadden:2010na,McFadden:2010vh,Bzowski:2011ab}.
See also
\cite{
Maldacena:2002vr,
Larsen:2002et,
Larsen:2003pf,Larsen:2004kf,vanderSchaar:2003sz,
Strominger:2001gp} for related studies.
Our focus here is how the improvement term changes the prediction of the observables.

The dS/CFT correspondence \cite{Strominger:2001pn} is based on the assumption that the wave function of an asymptotically de Sitter universe is equal to the partition function (or the Schwinger functional) of a three dimensional CFT,
\beq
\Psi_{\rm dS}[g]=Z_{\rm CFT}[g],
\eeq
where $g$ is collectively the three dimensional metric on the future boundary ${\F I}^+$ of the asymptotically de Sitter space as well as the scalar fields $g^I$, see Section \ref{sec:localRG}.
Using this wave function the expectation value of a field can be expressed e.g. as
$\langle f^2\rangle=\int{\C D}f f^2|\Psi_{\rm dS}[g]|^2$.
The duality implies that the boundary value of a field $f$ at ${\F I}^+$
acts also as the source of the corresponding operator ${\C O}$ in the CFT.
Following the assumption of dS/CFT, correlators of $f$ and those of ${\C O}$ are related, in momentum space, 
as
\cite{Maldacena:2002vr}
\begin{align}
\lish f_{\BS k}f_{-\BS k}\rish &=\frac{-1}{2 {\rm Re}\lish{\C O}_{\BS k}{\C O}_{-\BS k}\rish},
\cr
\lish f_{\BS k_1}f_{\BS k_2}f_{\BS k_3}\rish &
=\frac{2 {\rm Re}\lish{\C O}_{\BS k_1}{\C O}_{\BS k_2}{\C O}_{\BS k_3}\rish}
{\prod_i(-2 {\rm Re}\lish{\C O}_{\BS k_i}{\C O}_{-\BS k_i}\rish)}.
\label{eqn:dSCFTrel}
\end{align}
Note that $2 {\rm Re}$ originates from $|\Psi|^2$ in the expressions for the expectation values.
In cosmology, the metric fluctuations encodes information of the primordial density perturbations as well as of gravitational waves.
As the metric couples at the boundary to the energy-momentum tensor of the CFT,
the correlators of the metric fluctuations, and hence the cosmological parameters such as the
power spectrum of the density fluctuations, are obtained from the correlators of the energy-momentum tensor of the boundary theory.

The relations like \eqref{eqn:dSCFTrel} are also derived, without using the dS/CFT correspondence, 
but by first analytically continuing the asymptotically de Sitter background to 
the asymptotically AdS domain wall solution \cite{Boonstra:1998mp}, then applying the AdS/CFT duality and 
performing a second analytic continuation \cite{McFadden:2009fg,McFadden:2010na,McFadden:2010vh}.
A key feature in this approach is that the analytic continuation extends also to the
perturbation equations and hence the fluctuation spectra can be computed using the standard AdS/CFT dictionary.
In \cite{McFadden:2009fg,McFadden:2010na,McFadden:2010vh,Bzowski:2011ab} the two and three point correlators of the curvature perturbation 
are computed (in the regime where the gravity description is valid) as
\begin{align}
&\lish\zeta(k)\zeta(-k)\rish
=\frac{-1}{2{\rm Im}\lish T(q)T(-q)\rish},
\cr
&\lish\zeta(k_1)\zeta(k_2)\zeta(k_3)\rish\label{eqn:spectra}\\
&=-\frac{{\rm Im}\Big[
\lish T(q_1)T(q_2)T(q_3)\rish +\sum_i\lish T(q_i)T(-q_i)\rish-2\Big(
\lish T(q_1)\Upsilon (q_2,q_3)\rish + \mbox{cyclic perms}\Big)
\Big]}{4 {\rm Im}\lish T(q_1)T(-q_1)\rish {\rm Im}\lish T(q_2)T(-q_2)\rish 
{\rm Im}\lish T(q_3)T(-q_3)\rish},\nn
\end{align}
accompanied by analytic continuation of the momenta
$q=-ik$, 
$q_i=-ik_i$; $k, k_i\in{\B R}$.
Note that the semi-local contact terms are also included in the three-point correlator.
It is convenient to decompose the two point correlator of the energy-momentum tensor
into the trace part and the transverse-traceless part,
\begin{align}
\lish T_{ij}(q)T_{k\ell}(-q)\rish={\C A}(q)\Pi_{ijk\ell}+{\C B}(q)\pi_{ij}\pi_{k\ell},
\label{eqn:TijTkl_2}
\end{align}
where
\begin{align}
\Pi_{ijk\ell}=\frac{\pi_{ik}\pi_{j\ell}+\pi_{i\ell}\pi_{jk}-\pi_{ij}\pi_{k\ell}}{2},
\qquad
\pi_{ij}=\delta_{ij}-\frac{q_iq_j}{q^2},
\end{align}
are the projection operators.
The two point correlator of the trace of the energy-momentum tensor is then written as
$\lish T(q)T(-q)\rish=4{\C B}(q)$.
The power spectrum for the curvature perturbation is expressed using ${\C B}$ as
\begin{align}
\Delta_S^2(k)=\frac{k^3}{2\pi^2}\lish\zeta(k)\zeta(-k)\rish
=\frac{-k^3}{16\pi^2{\rm Im}{\C B}(q)}.
\label{eqn:PS_S}
\end{align}
The tensor mode $\gamma_{ij}$ of the metric perturbation couples to the 
transverse-traceless part of the energy-momentum tensor.
Its power spectrum is written using the 
${\C A}$ part of \eqref{eqn:TijTkl_2} as \cite{McFadden:2009fg}
\begin{align}
\Delta_T^2(k)=\frac{k^3}{2\pi^2}\lish \gamma_{ij}^*(k)\gamma^{ij}(-k)\rish
=\frac{-2k^3}{\pi^2{\rm Im}{\C A}(q)}.
\label{eqn:PS_T}
\end{align}

We consider a set of fields that are described by the three dimensional quantum 
field theory of Section \ref{sec:3dQFT}, and regard it as the holographic dual of the cosmological model.
Let us suppose that there are $\overline{N}_\xi$ bosonic fields having the same improvement parameter $\xi$.
The multiplicity $\overline{N}_\xi$ and the parameter $\xi$ characterize our holographic
model.
It is also possible to include various species of fields having different values of $\xi$, but we first focus for simplicity on the case where all the fields correspond to a single
$\xi$.
From \eqref{eqn:TijTkl_1} the values of ${\C A}(q)$ and ${\C B}(q)$, including the multiplicity of the fields, become
\begin{align}
{\C A}(q)=\frac{q^3}{256}\overline{N}_\xi,
\qquad
{\C B}(q)=\frac{q^3}{4}\overline{N}_\xi \left(\xi-\eighth\right)^2,
\end{align}
and \eqref{eqn:PS_S}, \eqref{eqn:PS_T} give the scalar and tensor power spectra of our model 
\begin{align}
\Delta_S^2(k)=\frac{16}{N_\xi\pi^2 (1-8\xi)^2},
\qquad
\Delta_T^2(k)=\frac{512}{N_\xi\pi^2}.
\label{eqn:PS}
\end{align}
Here, we have performed analytic continuation\footnote{
The model of the dual QFT considered in \cite{McFadden:2009fg,McFadden:2010na,McFadden:2010vh,Bzowski:2011ab} consists of ${\C N}_A$ gauge fields, ${\C N}_\phi$ minimally coupled scalars,
${\C N}_\chi$ conformally coupled scalars and ${\C N}_\psi$ fermions, all adjoint in
$SU(\overline{N})$.
The analytic continuation is made for the number of colors $N^2=-\overline{N}^2$.
The multiplicity in our model corresponds to 
$\overline{N}_\xi={\C N}_{A, \phi, \chi, \psi}\overline{N}^2$.}
$q=-ik$, $\overline{N}_\xi=-N_\xi$ and assumed $k\in{\B R}$.
This analytic continuation is necessary for the interpretation of the model using the domain wall solution.
Both spectra of \eqref{eqn:PS} are seen to be scale independent. 
In particular, we conclude that the density fluctuations have the Harrison-Zel’dovich
spectrum.
The spectrum can be shifted if loop corrections and/or deformations by nearly marginal operators are included \cite{Larsen:2003pf,Larsen:2004kf,Seery:2006tq,McFadden:2009fg,vanderSchaar:2003sz,Bzowski:2012ih}.
The amplitude of the scalar perturbation $\Delta_S^2(k)\sim 10^{-9}$ requires large
$N_\xi$.
The divergence of $\Delta_S^2(k)$ at $\xi=\eighth$ reflects the fact that in the conformal
limit the scalar fluctuations become pure gauge.
Observation\footnote{Detection of the CMB B-mode polarization 
was reported by the BICEP2 experiment \cite{Ade:2014xna}, with tensor/scalar ratio $r=0.20^{+0.07}_{-0.05}$.}
constrains the tensor-to-scalar ratio to be 
$r\approx 0.2$.
This is to be compared with
\begin{align}
r=\frac{\Delta_T^2(k)}{\Delta_S^2(k)}=32(1-8\xi)^2,
\end{align}
which indicates that $\xi$ is close to the conformal value, $|\xi-\eighth |\approx 10^{-2}$.

Using the results of Section \ref{sec:3dQFT} and the same analytic continuation as above,
the bispectrum in \eqref{eqn:spectra} is written as
\begin{align}
\lish\zeta(k_1)\zeta(k_2)\zeta(k_3)\rish
=\frac{5}{36}(24\xi+1)B^{\rm equil}_{\zeta}(k)-\frac{10}{9}\xi B^{\rm ortho}_{\zeta}(k),
\end{align}
where $B^{\rm equil}_{\zeta}(k)$ and $B^{\rm ortho}_{\zeta}(k)$ are the equilateral and 
orthogonal templates of the bispectrum (see Appendix \ref{sec:NG}). 
The nonlinearity parameters of the model are thus
\beq
f_{\rm NL}^{\rm local}=0,\quad
f_{\rm NL}^{\rm equil}=\frac{5}{36}(24\xi+1),\quad
f_{\rm NL}^{\rm ortho}=-\frac{10}{9}\xi.
\label{eqn:NG_pred}
\eeq
In view of the recent observational bound for these parameters
\cite{Ade:2013ydc} 
\beq
f_{\rm NL}^{\rm local}=2.7\pm 5.8,\quad
f_{\rm NL}^{\rm equil}=-42\pm 75,\quad
f_{\rm NL}^{\rm ortho}=-25\pm 39,
\label{eqn:NG_obs}
\eeq
our results \eqref{eqn:NG_pred} with $\xi\approx\eighth$ are safely within the bound but
do not provide significant constraints of the model in foreseeable future.
Furthermore, there are other sources of non-Gaussianities during the evolution of the
fluctuations, such as due to nonlinear interactions at superhorizon scales.
It is nevertheless worthwhile pointing out as a generic feature of the holographic cosmological model that the nonlinearity parameters are
intimately related with the parameter $\xi$ of the improvement.
It is intriguing to see that in the bispectrum two values of $\xi$ are special:
the minimal one $\xi = 0$ compatible with the shift symmetry makes the orthogonal component vanish, as observed in \cite{Bzowski:2011ab}; here we also find that
$\xi = -\frac{1}{24}$ makes the equilateral component vanish. 
The implication of the value $\xi = -\frac{1}{24}$ in the field theory in not
clear to us.

As alluded to above, this model can be generalized to have different values of the
improvement parameter $\xi_1, \xi_2, \cdots$.
Denoting the corresponding multiplicities of the scalar fields as $N_{\xi_1}$, $N_{\xi_2}$,
etc., the power spectra \eqref{eqn:PS} are modified as
\begin{align}
\Delta_S^2(k)
&=\frac{16}{\pi^2 \left[N_{\xi_1} (1-8\xi_1)^2+N_{\xi_2} (1-8\xi_2)^2+\cdots\right]},\cr
\Delta_T^2(k)
&=\frac{512}{\pi^2\left[N_{\xi_1}+N_{\xi_2}+\cdots\right]}.
\end{align}
The model studied in \cite{McFadden:2009fg,McFadden:2010na,McFadden:2010vh,Bzowski:2011ab} is a particular example of this type, with
$\xi=0$ for the gauge and the minimally coupled scalar fields and
$\xi=\eighth$ for the fermion and the conformally coupled scalar fields, which yields
$f_{\rm NL}^{\rm local}=0$,
$f_{\rm NL}^{\rm equil}=\frac{5}{36}$ and
$f_{\rm NL}^{\rm ortho}=0$.
\label{eqn:NG_obs}
It is crucial in their model that the gauge fields and the minimally coupled scalars
are not conformal and hence the quantum field theory is not conformal invariant.


\section{Discussions}
\label{sec:concl}

We have discussed the improvement ambiguity of the energy-momentum tensor
in the context of holographic cosmology.
One of the motivations for studying a holographic model of cosmology is the trans-Planckian
problem of inflationary cosmology, that is, the lack of reliable description in the UV region.
As our successful experience in string theory suggests that holography is a plausible nature of quantum gravity, 
it would be natural to suppose that the Universe in the very UV region may have an alternative
description in terms of a certain three dimensional quantum field theory.
In the absence of a guiding principle for finding the dual quantum field theory, 
the authors of \cite{
McFadden:2009fg,McFadden:2010na,McFadden:2010vh,Bzowski:2011ab} started investigating holographic cosmology using
gauge fields, fermions, and minimal and conformal scalars in three dimensions.
It seems to us, however, that there is no compelling reason to restrict ourselves to the minimal and conformal coupling of the fields.
In fact, in conformal field theory the energy-momentum tensor (or the Schwinger functional)
is fundamental, and it would be natural to start looking at properties of the energy-momentum tensor.

This naturally led us to consider a generalization of the model studied in \cite{
McFadden:2009fg,McFadden:2010na,McFadden:2010vh,Bzowski:2011ab}.
We computed the power spectrum and bispectrum of the curvature perturbations as well as
the tensor-to-scalar ratio.
The power spectrum is found to be the Harrison-Zel'dovich type, and the amplitude of the
power spectrum and the tensor-to-scalar ratio can be adjusted to agree with observations, 
by choosing appropriate values of the multiplicity of the fields $N_{\xi}$ and the improvement parameter $\xi$.
We also found that the bispectrum is generically a mixture of the equilateral and 
orthogonal shapes, while the local component is found to be zero.
In the light of the current observational sensitivity, the parameters
$f^{\rm equil}_{\rm NL}$ and $f^{\rm ortho}_{\rm NL}$
of non-Gaussianities, while non-zero, do not have
significance in foreseeable future.
We may also compute scalar-tensor three-point cross correlators, which are expected to be
even smaller.
The tensor three-point correlator, on the other hand, is coincidentally same as the particular case of \cite{Maldacena:2011nz}
since the improvement due to a scalar operator (i.e. $\int d^dx \sqrt{g} \delta\xi^\alpha RO_\alpha$) only affects the trace part of the energy-momentum tensor (see also \cite{Bzowski:2011ab}).\footnote{In most generality, this coincidence does not necessarily persist due to the lack of (manifest) conformal invariance. For instance the improvement of the form $\int d^dx \sqrt{g} \xi R_{ij} L^{ij}$ with traceless tensor operator $L^{ij}$ may change the observation made in \cite{Maldacena:2011nz}.}

In the main part of this paper, we have not discussed deformation of the theory that drives the renormalization group flow (see Appendix A).
This corresponds to nontrivial $\beta$-functions appearing in the trace identity \eqref{traceid}. 
It would certainly be interesting to discuss the model including such a term following \cite{Larsen:2003pf,Larsen:2004kf,Seery:2006tq,vanderSchaar:2003sz,Bzowski:2012ih}, in particular, for discussing the holographic cosmological model in the light of
precision measurements.

In the model discussed above the conformal invariance is broken by the improvement term of the energy-momentum tensor.
From the holographic viewpoint this is related to the boundary conditions for the scalar field which mixes the boundary metric and the boundary scalar sources. 
In realistic cosmological model building, it would be crucial to discuss which boundary condition is realized.
This is directly related to the problem of the transition mechanism to the radiation dominated era, as the metric perturbations are to be eventually converted into the matter fluctuations of the Universe.
While this topic is beyond the scope of the present paper, it would be certainly an important question to be answered in the future with hopefully better understanding of holographic cosmology, since our prediction relies on it.

{\bf Acknowledgements}\\
We acknowledge helpful conversations with Gary Shiu and Toshifumi Noumi.
S.K. acknowledges warm hospitality of the Kavli IPMU, the University of Tokyo where the final stage of this work was carried out. He also thanks Paul McFadden for sharing unpublished lecture notes.
This work is supported in part by the Sherman Fairchild Senior Research Fellowship at California Institute of Technology with DOE grant DE-FG02-92ER40701 and the World Premier International Research Center Initiative (WPI Initiative), MEXT, Japan (Y.N.),
and by the National Research Foundation of Korea Grant-in-Aid for Scientific Research 
No. 2013028565 (S.K.).

\appendix

\section{Local renormalization group equations}

In the main part of this paper, we focused mostly on the (A)dS/CFT correspondence  at the (conformal) fixed point.
To study a more realistic model of holographic cosmology, it will be vital to go beyond the fixed point, for the following two reasons. Firstly, the interpretation as the slow roll inflation means that the de Sitter invariance must be broken due to the slow running of the inflaton, or by the source term of the  corresponding dual operator in the dual picture. 
The corresponding physics is not transparent in the strongly coupled gravity regime, but at least we will need a (holographic) mechanism to end inflation.
The observationally supported tilt of the spectrum $n_s\approx 0.96$ should also be accounted for by inclusion of such an operator.\footnote{
The scalar power spectrum may be tilted alternatively by the improvement term.
If the energy-momentum tensor is improved by an operator ${O_\alpha}$ having dimension 
$\Delta_{O_{\alpha}} =d-2 +\lambda$ with non-zero $\lambda$, then the spectral index becomes
$n_s-1 = -2\lambda$. Note that this does not affect the scale invariance of the dual field theory as 
the improvement part of the energy-momentum tensor does not need to have dimension $d$. In local renormalization group, the difference of the dimension from $d-2$ is encoded by the anomalous dimension $\gamma^{\alpha}_{\ \beta}$ in \eqref{dilatation}.
Tilt of the tensor mode $n_t$ (if observed) cannot be obtained in this way.}
Secondly, for application of the improvement ambiguities in the energy-momentum tensor, we have to determine which energy-momentum tensor is coupled with our particle physics (and dark matter) as discussed in section \ref{sec:localRG}.

The change of the scalar fields under the conformal time in bulk geometry corresponds to the running of the scalar coupling constant in the dual field theory. With some technical assumptions in holography,
the bulk equations of motion can be interpreted from the local renormalization group equations through holographic renormalization group.\footnote{The relation is expected but slightly non-trivial. For instance, the bulk equations of motion are typically second order while the renormalization group equation is first order. The first principle derivation of the holographic bulk equations from local renormalization group in large $N$ limit can be found in \cite{Lee:2013dln}\cite{Nakayama:2014cca}.} With the correspondence in mind, in this appendix we would like to study the local renormalization group equation relevant for our improvement ambiguities in the energy-momentum tensor.

Let us consider the Schwinger functional
\begin{align}
e^{-W[g_{ij}(x), g^I(x), m_{\alpha}(x)]} = \int \mathcal{D}X e^{-S[X] - \int d^3x\sqrt{g} g^I(x) O_I(x) + m^\alpha(x) O_{\alpha}(x)} .
\end{align}
For simplicity, we will not consider the vector operators.
In the power-counting renormalization scheme, the Schwinger functional satisfies the local renormalization group equation \cite{Osborn:1991gm}
\begin{align}
\Delta_{\sigma} W = A_{\sigma}  , \label{localrg}
\end{align}
where $\Delta_{\sigma}$ is the local renormalization group operator
\begin{align}
\Delta_{\sigma} &= \int d^3x \sqrt{g} \left( 2\sigma g_{ij} \frac{\delta}{\delta g_{ij}} + \sigma \beta^I \frac{\delta}{\delta g^I}  \right) + \Delta_{\sigma,m} \cr
\Delta_{\sigma,m} &= -\int d^3x \sqrt{g} \left( (\sigma(2-\gamma)^\alpha_{\ \beta}m^\beta + \frac{1}{4}\sigma R \eta^\alpha  + \sigma \delta_{I}^\alpha (\Box g^I) + \sigma \epsilon_{IJ}^\alpha \partial^i g^I \partial_i g^J \right.   \cr
 & \left. \left. + 2\partial_i \sigma (\theta_I^\alpha \partial^i g^I) + (\Box \sigma) \kappa^\alpha \right)  \frac{\delta}{\delta m^\alpha} \right) \ \label{dilatation}
\end{align}
and $A_{\sigma}$ is the Weyl anomaly that is a local functional constructed out of $g_{ij}(x)$, $g^I(x)$ and $m^{\alpha}(x)$. The Weyl anomaly in $d=3$ within the local renormalization group was studied in \cite{Nakayama:2013wda}, but we will not use them in the following.

Neglecting the anomaly in the flat spacetime limit, the local renormalization group equation \eqref{localrg} is equivalent to the trace identity \eqref{traceid}. On the other hand, the physical meaning of \eqref{localrg} in the bulk via holography is nothing but the Hamiltonian constraint: The wave function of the universe is annihilated (up to anomaly) by the Hamiltonian density operator
\begin{align}
\mathcal{H}_{\sigma} \Psi[g_{ij}(x), g^I(x), m^\alpha(x)] = 0, 
\end{align}
which is the consequence of the time reparametrization invariance.\footnote{There is one cautious remark on the meaning of ``anomaly". The anomaly term $A_{\sigma}$ in the local renormalization group can be regarded as the local renormalization of the cosmological constant operator. Then, the local renormalization group equation is anomaly-free \cite{Nakayama:2013wda}. This makes the Hamiltonian constraint anomaly free as it should. Indeed, the ``anomaly" term plays a crucial role in deriving the bulk equations of motion in holography \cite{Nakayama:2014cca}.}
Intuitively this is the reason why we identify the running of the coupling constant in the local renormalization group as the roll of inflaton in the slow roll approximation. The neglect of the second order nature of the bulk equations of motion is justified by the slow roll assumption. The fluctuation of the scalar mode of the metric is related to the fluctuation of the inflaton through the trace identity \eqref{traceid}. This is essentially the origin of the ``inflation consistency conditions" from holography \cite{Creminelli:2012ed}\cite{Schalm:2012pi}.\footnote{There is also a less important ``momentum constraint" that is obtained from requiring that the $d$-dimensional diffeomorphism is preserved at each renormalization scale.}

Now we would like to study the ambiguities of the energy-momentum tensor from the viewpoint of the ambiguities in the local renormalization group \cite{Nakayama:2013wda}. The Schwinger functional $W$ (identified with the GKP-W partition functional) is ambiguous with respect to the variation of the local renormalization group transformation generator $\Delta_{\sigma}$ 
\begin{align}
\delta \Delta_\sigma = [\mathcal{D},\Delta_{\sigma}] \cr
\delta W = \mathcal{D} W  , \label{class2a}
\end{align}
where 
\begin{align}
\mathcal{D} = \int d^3x \sqrt{g} \left(\frac{1}{4}R h^\alpha + (\Box g^I) d^\alpha_I + (\partial_i g^I \partial^i g^J) e^\alpha_{IJ} \right) \frac{\delta}{\delta m^\alpha} . 
\end{align}
Accordingly, the local renormalization functions change as
\begin{align}
\delta \eta^\alpha &= \beta^I \partial_I h^\alpha - \gamma_{\ \beta}^{\alpha} h^\beta \cr
\delta \kappa^\alpha &= - h^\alpha + d^\alpha_I\beta^I \cr
\delta \theta_I^\alpha &= \frac{1}{2} d^\alpha _I + (\partial_I \beta^J) d^\alpha_J + e^\alpha_{IJ} \beta^J \cr
\delta \delta^\alpha_I &= \mathcal{L}_\beta d_I^\alpha  - \gamma_{\ \beta}^\alpha d^\beta_I \cr
\delta \epsilon^\alpha_{IJ} &=  \mathcal{L}_\beta e_{IJ}^\alpha - \gamma_{\ \beta}^\alpha e_{IJ}^\beta + 
(\partial_I \partial_J \beta^K)d_K^\alpha  ,
\end{align}
where $\mathcal{L}_\beta$ is the Lie derivative with respect to the vector field $\beta^I$ in coupling constant space parametrised by $g^I$. In holography, the class 2 ambiguity corresponds to the field redefinition in the bulk, the simplest example of which was discussed in section 3 of the main text.

In relation to our applications to holographic cosmology, we note the following two things. First of all by using the renormalization group ambiguities, we may always make $\kappa^\alpha$ vanish by choosing appropriate $h^\alpha$.
The second point is the energy-momentum tensor is renormalized due to the $\eta^\alpha$ term in \eqref{dilatation}. Under the renormalization group flow, the improvement term is renormalized as
\begin{align}
\frac{d}{d\log\mu} T_{ij} =-\frac{1}{2}(\partial_i \partial_j - \Box \delta_{ij})  \eta^\alpha O_{\alpha} \label{renormalem} . 
\end{align}
In general, we may not be able to set $\eta^\alpha = 0$, and we should keep track of which operator couples with the (background) gravity along the renormalization group flow. 
We should emphasize here that  a priori there is no natural choice of the energy-momentum tensor within the local renormalization group equation, and we need the other physical input to fix the ambiguity.

The local renormalization functions are, however, not arbitrary. The Wess-Zumino consistency condition $[\Delta_{\sigma}, \Delta_{\tilde{\sigma}}] = 0$ from the Abelian nature of the local renormalization group transformation demands
\begin{align}
\eta^\alpha &= \delta^\alpha_I \beta^I - (\beta^I \partial_I \kappa^\alpha - \gamma_{\ \beta}^\alpha \kappa^\beta) \cr
\delta_I^\alpha + 2(\partial_I \beta^J) \delta^\alpha_J + 2 \epsilon^\alpha_{IJ} \beta^J &= 2{\mathcal{L}}_{\beta}\theta_I^\alpha - 2\gamma_{\ \beta}^\alpha \theta_I^\beta .  \label{consis}
\end{align}
These conditions should be understood as the consistency of the Hamiltonian constraint in the dual gravitational system from holography. At the abstract level, the condition $[\mathcal{H}_{\sigma}, \mathcal{H}_{\tilde{\sigma}}] = 0$ should follow trivially from the time reparametrization invariance manifested in $d+1$-dimensional action, but the direct check of the commutator (or Poisson bracket in the classical limit) requires  a detailed knowledge of the dynamics from $d$ dimensional viewpoint.

Finally, we would like to mention one additional field redefinition ambiguity of the Schwinger functional, which we do not typically consider in the dual field theory side but may be more common in the holographic side. Take
\begin{align}
\mathcal{D} = \int d^d x \sqrt{g} \left( f(g^I) g_{ij} \frac{\delta}{\delta g_{ij}} \right) \ 
\end{align}
in the above class 2 ambiguity \eqref{class2a}.
This induces the shift of the local renormalization group operator $\Delta_{\sigma}$ by the amount
\begin{align}
\int d^d x \sqrt{g} \sigma (\beta^I \partial_If) g_{ij} \frac{\delta}{\delta g_{ij}} .
\end{align}
The effect is only shifting the overall normalization of the energy-momentum tensor, and what we mean by the scale under the local renormalization group transformation. 
Alternatively, one may think that we have made a finite wave function renormalization for the composite operator $T^{i}_{\ i}$.

On the other hand, the field redefinition of the metric with respect to (nearly) massless fields
\begin{align}
g_{\mu\nu} \to f(\Phi^I) g_{\mu\nu}
\end{align}
 is popularly used in the gravity side in order to set the canonically normalised Einstein frame. In relation to the holographic cosmology, it is again important to choose the most natural ``metric" from the viewpoint of our particle physics action. The prediction of the cosmological fluctuation does depend on the assumption how the particle physics sector couples to the inflaton sectors. A different choice would give additional contributions such as curvaton mechanism. While we will focus on the universal energy-momentum tensor contributions in our sample computation in the main text, we should keep these possible additional contributions from the light scalars in mind.\footnote{When they couple differently to different matters in our particle physics sector, they induce the isocurvature fluctuation.}  Of course, once we fix how the light fields couple to our particle physics after inflation, the choice of the frame is just a gauge choice, and it cannot affect the final outcome as reviewed e.g. in \cite{Yamaguchi:2011kg}. From the holographic renormalization group viewpoint, this is the scheme (in)dependence of the renormalization group. As long as we have $g^I$ fixed at the future boundary, our prediction of the power spectra does not change in the physical unit (e.g. in Planck scale which we have set unity) since the overall normalization of the energy-momentum tensor is not physical.

\section{Correlators in the dual field theory}
\label{sec:FTcomp}

In this appendix we delineate the derivation of the correlation functions
\eqref{eqn:TijTkl_1},\eqref{eqn:TTT},\eqref{eqn:TUpsilon} which appeared in Section
\ref{sec:3dQFT}.
The derivation is somewhat tedious but entirely straightforward.
These correlation functions were used in Section \ref{sec:observables} to find the spectra of the fluctuations in our model of holographic cosmology.

We first note that the energy-momentum tensor (\ref{eqn:Tij1}) may be written as
\beq
T_{ij}(x)=(1-2\xi) :\partial_i\phi\partial_j\phi:
+(2\xi-\half)\delta_{ij}:(\partial\phi)^2:-2\xi :\phi\partial_i\partial_j\phi:
+2\xi \delta_{ij}:\phi\Box\phi:,
\label{eqn:Tij2}
\eeq
in the flat space-time limit,
where the normal ordering is now indicated explicitly. 
Each term represents a dimension three operator.
The (cross-) correlation functions of these operators are found by direct free field
computation using Feynman integrals.
After dimensional regularization (see e.g. \cite{Leibbrandt:1975dj}) we find, in the 
momentum space,
\bea
&&\lish :\partial_i\phi\partial_j\phi :: \partial_k\phi\partial_\ell\phi: \rish({\BS q})
=\Big\{
3q_i q_j q_k q_\ell +\big[
3(\delta_{ij}q_k q_\ell 
+\delta_{k\ell}q_i q_j)
-\delta_{jk}q_i q_\ell
-\delta_{i\ell}q_j q_k\nn\\
&&\qquad\qquad -\delta_{ik}q_j q_\ell 
-\delta_{j\ell}q_i q_k
\big]q^2 
+(\delta_{ij}\delta_{k\ell}+\delta_{i\ell}\delta_{jk}+\delta_{ik}\delta_{j\ell})q^4
\Big\}\frac{1}{512 q},\label{eqn:dphdphdphdph}\\
&&\lish :\partial_i\phi\partial_j\phi :: \phi\partial_k\partial_\ell\phi: \rish({\BS q})
=\Big\{
5q_i q_j q_k q_\ell +\left[
5\delta_{ij}q_k q_\ell 
-3\delta_{k\ell}q_i q_j
+\delta_{jk}q_i q_\ell
+\delta_{i\ell}q_j q_k
\right.\nn\\
&&\qquad\qquad\left.
+\delta_{ik}q_j q_\ell 
+\delta_{j\ell}q_i q_k
\right]q^2 
-(\delta_{ij}\delta_{k\ell}+\delta_{i\ell}\delta_{jk}+\delta_{ik}\delta_{j\ell})q^4
\Big\}\frac{1}{512 q},\label{eqn:dphdphphddph}\\
&&\lish :\phi\partial_i\partial_j\phi :: \partial_k\phi\partial_\ell\phi: \rish({\BS q})
=\Big\{
5q_i q_j q_k q_\ell +\left[
-3\delta_{ij}q_k q_\ell 
+5\delta_{k\ell}q_i q_j
+\delta_{jk}q_i q_\ell
+\delta_{i\ell}q_j q_k
\right.\nn\\
&&\qquad\qquad\left.
+\delta_{ik}q_j q_\ell 
+\delta_{j\ell}q_i q_k
\right]q^2 
-(\delta_{ij}\delta_{k\ell}+\delta_{i\ell}\delta_{jk}+\delta_{ik}\delta_{j\ell})q^4
\Big\}\frac{1}{512 q},\label{eqn:phddphdphdph}\\
&&\lish :\phi\partial_i\partial_j\phi :: \phi\partial_k\partial_\ell\phi: \rish({\BS q})
=\Big\{
19q_i q_j q_k q_\ell -\left[
5\delta_{ij}q_k q_\ell
+5\delta_{k\ell}q_i q_j
+\delta_{jk}q_i q_\ell
+\delta_{i\ell}q_j q_k
\right.\nn\\
&&\qquad\qquad\left.
+\delta_{ik}q_j q_\ell 
+\delta_{j\ell}q_i q_k
\right]q^2 
+(\delta_{ij}\delta_{k\ell}+\delta_{i\ell}\delta_{jk}+\delta_{ik}\delta_{j\ell})q^4
\Big\}\frac{1}{512 q}.\label{eqn:phddphphddph}
\eea
The double chevrons $\lish\cdots\rish$ indicate omission of the $(2\pi)^3\delta(\Sigma q_i)$ 
factor and $q\equiv|{\BS q}|$, as noted in the main text.
%
%
%
%
Combining these, we find $\lish T_{ij}T_{k\ell}\rish (q)$ in the form of
\eqref{eqn:TijTkl_1}.


Computation of the bispectrum requires the three point correlator of the trace of the
energy-momentum tensor $\lish TTT\rish$ as well as the two point correlator 
$\lish T\Upsilon\rish$, where $\Upsilon$ is the semilocal operator defined in (\ref{eqn:Upsilon}).
Free field computation gives 
\bea
&&\lish :(\partial\phi)^2 (q_1)::(\partial\phi)^2 (q_2)::(\partial\phi)^2 (q_3):\rish
=-\eighth q_1q_2q_3,\\
&&\lish :(\partial\phi)^2 (q_1)::(\partial\phi)^2 (q_2)::\phi\Box\phi (q_3):\rish
=\frac{1}{16}q_3(q_1^2+q_2^2-q_3^2),\\
&&\lish :(\partial\phi)^2 (q_1)::\phi\Box\phi (q_2)::\phi\Box\phi (q_3):\rish
=\frac{1}{16} q_1(q_1^2-q_2^2-q_3^2),\\
&&\lish :\phi\Box\phi (q_1)::\phi\Box\phi (q_2)::\phi\Box\phi (q_3):\rish=0,
\eea
and
\bea
\lish :(\partial\phi)^2 (q_1):\Upsilon(q_2,q_3)\rish
&=&\frac{\xi}{16} q_1(q_1^2-q_2^2-q_3^2),\\
\lish :\phi\Box\phi(q_1): \Upsilon(q_2,q_3)\rish
&=&0,
\eea
where $q_1+q_2+q_3=0$ arising from the momentum conservation has been used.
We obtain \eqref{eqn:TTT} and \eqref{eqn:TUpsilon} from the formulae above.

\section{Non-Gaussianities}
\label{sec:NG}

In this appendix we summarize our conventions of the scalar power spectrum and the 
bispectrum.
We essentially follow \cite{Komatsu:2010fb,Bennett:2012zja,Ade:2013ydc}.

The two point correlation function of the primordial curvature perturbation $\zeta({\BS x})$ is written in the momentum space as
$\langle\zeta_{{\BS k}_1}\zeta_{{\BS k}_2}\rangle
=(2\pi)^3\delta^3\left({\BS k}_1+{\BS k}_2\right)
\lish\zeta(k_1)\zeta(-k_1)\rish$.
The delta function enforces momentum conservation resulting from the translational invariance.
Due to rotational invariance
$\lish \zeta(k)\zeta(-k)\rish$ depends only on the magnitude of the momenta 
$k=\vert{\BS k}\vert$.
The power spectrum of the curvature perturbations is conventionally defined by
\beq
\Delta_S^2(k)=\frac{k^3}{2\pi^2}\lish\zeta(k)\zeta(-k)\rish.
\eeq

The angular bispectrum of the curvature perturbations $B_\zeta$ is defined by the 3-point function as
\beq
\langle\zeta_{{\BS k}_1}\zeta_{{\BS k}_2}\zeta_{{\BS k}_3}\rangle 
=(2\pi)^3\delta^3\left({\BS k}_1+{\BS k}_2+{\BS k}_3\right)B_{\zeta}(k_1,k_2,k_3).
\eeq
Hence in our notation, $B_\zeta (k_1,k_2,k_3)=\lish\zeta(k_1)\zeta(k_2)\zeta(k_3)\rish$.
The translational invariance enforces the three momenta to add up to zero and form a triangle, 
with the three sides having lengths $k_1$, $k_2$, $k_3$.
Due to rotational invariance the bispectrum depends only on the magnitude of the momenta
$k_i=\vert{\BS k_i}\vert$.
Different shapes of the triangle correspond to different profiles of non-Gaussianities.
Commonly used templates are the local bispectrum \cite{Gangui:1993tt,Komatsu:2001rj}
\begin{align}
B_\zeta^{\rm local}({k}_1, {k}_2, {k}_3)&=\frac 65\Big\{P_\zeta({k}_1)P_\zeta({k}_2)
+P_\zeta({k}_2)P_\zeta({k}_3)+P_\zeta({k}_3)P_\zeta({k}_1)\Big\}\cr
&= \frac 65 (2\pi^2\Delta_S^2)^2
\left(
\frac{1}{k_1^3k_2^3}+\frac{1}{k_2^3k_3^3}+\frac{1}{k_3^3k_1^3}\right),
\label{eqn:Blocal}
\end{align}
the equilateral bispectrum \cite{Creminelli:2005hu}
\begin{align}
B_\zeta^{\rm equil}({k}_1, {k}_2, {k}_3)
=&\frac{18}{5}\Big\{
-P_\zeta({k}_1)P_\zeta({k}_2)-P_\zeta({k}_2)P_\zeta({k}_3)-P_\zeta({k}_3)P_\zeta({k}_1)\cr
&-2P_\zeta({k}_1)^{\frac 23}P_\zeta({k}_2)^{\frac 23}P_\zeta({k}_2)^{\frac 23}
+\left[
P_\zeta({k}_1)P_\zeta({k}_2)^{\frac 23}P_\zeta({k}_3)^\third
+\mbox{5 perms}\right]\Big\},\cr
=&\frac{18}{5}(2\pi^2\Delta_S^2)^2
\Big(
-\frac{1}{k_1^3k_2^3}-\frac{1}{k_2^3k_3^3}-\frac{1}{k_3^3k_1^3}-\frac{2}{k_1^2k_2^2k_3^2}\cr
&+\frac{1}{k_1k_2^2k_3^3}+\frac{1}{k_1k_2^3k_3^2}+\frac{1}{k_1^2k_2k_3^3}+\frac{1}{k_1^2k_2^3k_3}+\frac{1}{k_1^3k_2k_3^2}+\frac{1}{k_1^3k_2^2k_3}\Big),
\label{eqn:Bequil}
\end{align}
and the orthogonal bispectrum \cite{Senatore:2009gt}
\begin{align}
B_\zeta^{\rm ortho}({k}_1, {k}_2, {k}_3)
=&\frac{18}{5}\Big\{
-3\big[P_\zeta({k}_1)P_\zeta({k}_2)+P_\zeta({k}_2)P_\zeta({k}_3)+P_\zeta({k}_3)P_\zeta({k}_1)\big]\cr
&-8P_\zeta({k}_1)^{\frac 23}P_\zeta({k}_2)^{\frac 23}P_\zeta({k}_2)^{\frac 23}
+3\left[P_\zeta({k}_1)P_\zeta({k}_2)^{\frac 23}P_\zeta({k}_3)^\third+\mbox{5 perms}\right]\Big\}\cr
=&\frac{18}{5}(2\pi^2\Delta_S^2)^2
\Big(-3\left[
\frac{1}{k_1^3k_2^3}+\frac{1}{k_2^3k_3^3}+\frac{1}{k_3^3k_1^3}\right]
-\frac{8}{k_1^2k_2^2k_3^2}\cr
&+3\left[\frac{1}{k_1k_2^2k_3^3}+\frac{1}{k_1k_2^3k_3^2}+\frac{1}{k_1^2k_2k_3^3}+\frac{1}{k_1^2k_2^3k_3}+\frac{1}{k_1^3k_2k_3^2}+\frac{1}{k_1^3k_2^2k_3}\right]\Big),
\label{eqn:Bortho}
\end{align}
where $P_\zeta(k)\equiv\lish\zeta(k)\zeta(-k)\rish$.
The nonlinearity parameters are defined as the coefficients of these templates,
\beq
B_{\zeta}(k_1,k_2,k_3)
=f_{\rm NL}^{\rm local}B_\zeta^{\rm local}({k}_1, {k}_2, {k}_3)
+f_{\rm NL}^{\rm equil}B_\zeta^{\rm equil}({k}_1, {k}_2, {k}_3)
+f_{\rm NL}^{\rm ortho}B_\zeta^{\rm ortho}({k}_1, {k}_2, {k}_3).
\eeq

\input{virialc_4.bbl}

\end{document}

%% file: virialc_4.bbl
\providecommand{\href}[2]{#2}\begingroup\raggedright\endgroup